# From Participatory Sensing to Mobile Crowd Sensing

Bin Guo, Zhiwen Yu, Xingshe Zhou
School of Computer Science
Northwestern Polytechnical University
Xi'an, P. R. China
guobin.keio@gmail.com

Daqing Zhang
Telecommunication Network & Services Department
Institut TELECOM SudParis
Evry Cedex, France
daqing.zhang@it-sudparis.eu

*Abstract*—The research on the efforts of combining human and machine intelligence has a long history. With the development of mobile sensing and mobile Internet techniques, a new sensing paradigm called Mobile Crowd Sensing (MCS), which leverages the power of citizens for large-scale sensing has become popular in recent years. As an evolution of participatory sensing, MCS has two unique features: (1) it involves both implicit and explicit participation; (2) MCS collects data from two user-participant data sources: mobile social networks and mobile sensing. This paper presents the literary history of MCS and its unique issues. A reference framework for MCS systems is also proposed. We further clarify the potential fusion of human and machine intelligence in MCS. Finally, we discuss the future research trends as well as our efforts to MCS.

## I. INTRODUCTION

Successful society and city management relies on urban and community dynamics monitoring to provide essential information for decision making. In traditional sensing techniques such as wireless sensor networks (WSNs), distributed sensors are leveraged to acquire real-world conditions. There has been a growing body of studies on WSNs, however, commercial sensor network techniques have never been successfully deployed in the real world due to several reasons, such as insufficient node coverage, high installation/maintenance cost, and lack of scalability.

**Mobile Crowd Sensing (MCS)** presents a new sensing paradigm based on the power of mobile devices. The sheer number of user-companioned devices, including mobile phones, wearable devices, and smart vehicles, and their inherent mobility enables a new and fast-growing sensing paradigm: *the ability to acquire local knowledge through sensor-enhanced mobile devices – e.g., location, personal and surrounding context, noise level, traffic conditions, and in the future more specialized information such as pollution – and the possibility to share this knowledge within the social sphere, healthcare providers, and utility providers*. The information collected on the ground combined with the support of the cloud where data fusion takes place, make MCS a versatile platform that can often replace static sensing infrastructures, and enabling a broad range of applications including urban dynamics mining, public safety, traffic planning, environment monitoring, just to name a few.

A formal definition of MCS is: *a new sensing paradigm that empowers ordinary citizens to contribute data sensed or generated from their mobile devices, aggregates and fuses the data in the cloud for crowd intelligence extraction and people-centric service delivery*. Similar to participatory sensing [1], MCS has sensed data from mobile devices as inputs. Nevertheless, it additionally "reuses" the user-contributed data from mobile Internet services (mostly mobile social network services), which is often termed as the participatory media [2]. In other words, the user-contributed data are used for a second purpose. MCS further explores the integration and fusion of the data from heterogeneous, cross-space data sources. We use the following example to showcase its characters.

*The Urban Sensing Scenario*. Route planning [3] is a common type of application of MCS. With participatory sensing, we can collect GPS trajectory data from vehicles and compute the optimal route when answering a query with departure and destination points. However, for a more complex query, that is, to generate an itinerary for a visitor to a city given the time budget (start time, end time). It is not possible to leverage the single trajectory dataset. Further information such as POIs in the city, categories of each POI, the best time to visit the POIs, user preferences to different types of POIs, are further needed. These information can be obtained by reusing the user-contributed data from a mobile social network service (e.g., FourSquare). A similar example is noise mapping [4], which is also a popular type of MCS application. With participatory sensing, we can get the noise map using mobile audio sensing. But people may wonder the causes of noise in a specific place, which often correlates with the category (e.g., market, school, street) of that place. This, however, can be obtained from a LBSN check-in dataset. Therefore, with MCS, we can leverage both online and offline data contributed by participants and explore cross-space data fusion to nurture novel applications.

Numerous and unique research challenges arise from the mobile crowd sensing paradigm, ranging from styles of data collection, proper incentive mechanisms, quality of user-contributed data, cross-space data fusion, and so on. Further, MCS essentially represents a hybrid of human and machine intelligence, which has been little explored before. The key contribution of this paper can be summarized as follows:

- Give a literature history and the definition of MCS, and explain the evolution from participatory sensing to MCS;
- Clarify the technical foundation of MCS—the optimal fusion of human and machine intelligence, and present the key research issues of MCS;
- Propose a reference framework of MCS and discuss the future research trends.

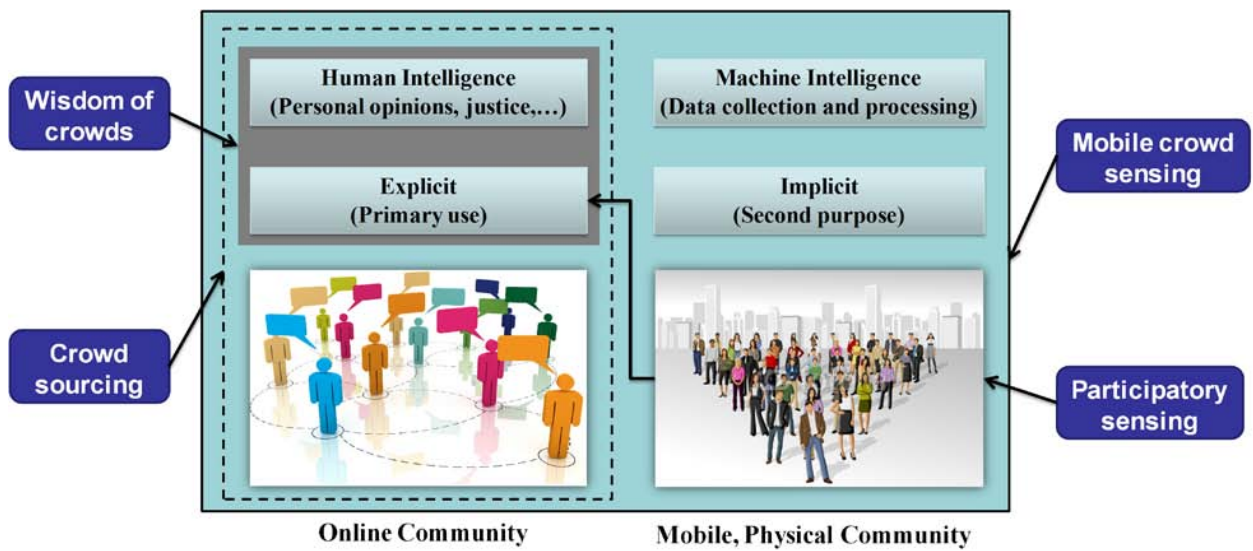

Figure 1. A comparison of MCS and related concepts.

## II. FROM PARTICIPATORY SENSING TO MOBILE CROWD SENSING

From the AI perspective, MCS is founded on a distributed problem-solving model where a crowd is engaged in solving a complex problem through an open call [5]. In the literature history, the concept of crowd-powered problem-solving has been explored in several research areas. One decade ago, Surowiecki has written a book titled "*The Wisdom of Crowds*" (or crowd wisdom) [6], where a general phenomena ─ the aggregation of information in groups, resulting in decisions that are often better than could have been made by any single member of the group─is revealed. It identifies four key qualities that make a crowd smart: *diversity in opinion*, *independence of thinking*, *decentralization*, and *opinion aggregation*. A similar concept to crowd wisdom is "*collective intelligence*" [7]. Different from the two concepts that focus on the advantages of group decision making, MCS is mainly about the crowd-powered data collection and analyzing process.

In 2005, two senior editors from Wired Magazine, Jeff Howeand and Mark Robinson, coined the term "*crowdsourcing*". According to the Merriam-Webster Dictionary[1], crowdsourcing is defined as *the practice of obtaining needed services or content by soliciting contributions from a large group of people, and especially from an online community*. A typical example is Wikipedia, where thousands of contributors from across the world have collectively created the world's largest encyclopedia. Compared to MCS, crowdsourcing focuses on the participation of online crowds.

The most close concept to MCS is *participatory sensing*, proposed by Burke et al. in 2006 [1]. It tasks everyday mobile devices to form interactive, participatory sensor networks that enable public and professional users to gather, analyze and share local knowledge. The definition of participatory sensing emphasizes explicit user participation when it was proposed. In recent years, with the development of *mobile sensing* and *mobile Internet* techniques, the scope of crowd problem-solving systems using mobile devices has been broadened. To this end, we extend the definition of participatory sensing from two aspects and term the new concept *mobile crowd sensing* (MCS). First, MCS leverages both sensed data from mobile devices (from physical community) and user-contributed data from mobile social network services (from online community). Second, MCS counts both explicit and implicit user participation (details will be clarified in the next section).

A comparison of MCS and related areas is given in Fig. 1. We can find that both crowd wisdom and crowdsourcing rely on human intelligence; while participatory sensing and MCS explore a fusion of human and machine intelligence (we discuss this later in Section V). Compared to participatory sensing, MCS can have both explicit (primary purpose) and implicit (second purpose) user participation and allow cross-space (online& offline) data fusion.

## III. KEY FEATURES OF MCS

Mobile crowd sensing opens a new sensing paradigm for crowd-powered sensing. This section characterizes its key features.

### A. Citizen Participation: Explicit or Implicit

The involvement of citizens in the sensing loop is the chief feature of participatory sensing. The same is true for MCS, but it moves a step further than participatory sensing.

*Two data generation modes*. Varied human-companioned devices can act as mobile sensor nodes in MCS, including mobile phones, wearable devices, smart vehicles, smart cards, and so on. There are two different data generation modes in MCS: *mobile sensing* and *user-generated data in mobile social network services* (e.g., LBSNs).

- *Mobile sensing*. It typically functions at a context-based and individual manner, leveraging the rich sensing capabilities from individual devices. It is the data collection method used by participatory sensing.

---

[1] http://www.merriam-webster.com/dictionary/crowdsourcing

- *Mobile social networks.* With the rapid development of mobile Internet, mobile social network services that bridge the gap between online interactions and physical elements (e.g., check-in places, activities [8], objects [9]) are fast growing. The large-scale user-contributed data opens a new window to understand the dynamics of the city and society, which is counted as the other data source for mobile crowd sensing.

In summary, the combination of participatory mobile sensing and participatory mobile social network data is a unique feature of MCS.

*The sensing style.* Previous studies often discuss about user-participated sensing at one dimension: *the degree of user involvement in the sensing process*. As presented by Ganti et al. [10], crowd-powered sensing can span a wide spectrum in terms of user involvement, with participatory and opportunistic sensing at the two ends. The proportion of human involvement will depend on application requirements and device capabilities. With the two data generation modes in MCS, we intend to category the sensing style from a new dimension: *the awareness of participants to the sensing task.* For both participatory and opportunistic sensing, data collection is the primary-purpose of the application. The sensing task is therefore *explicit* to the user (as she is informed). For user-contributed data from MSN services, however, the data is used for a second purpose (the primary purpose is online social interaction), and thus it is performed in an *implicit* manner. We thus characterize the sensing style of MCS at two levels: *explicit* and *implicit*.

The involvement of human participation in the computing process will lead to a mixture of human and machine intelligence in MCS. How to optimize the combination of human and machine intelligence, becomes a significant design issue for MCS systems.

### B. User Motivation

In data sharing among peers, the development of a solid economic model is highly important. This issue is even more critical when the devices (e.g., mobile phones, wearable sensors) have very limited resources (e.g., energy and storage capacity) or the information revealed is highly sensitive. Deploying a mobile crowd sensing system on a wide scale requires a large number of participants. Participants may drop out of the collecting loop unless return on investment is greater than their expectations. Questions about human motivation have been central in philosophy and economics. The promise of *financial* or *monetary* gain is an important incentive method for most actors in markets and traditional organizations. *Interest* and *entertainment* are also important motivators in many situations, even when there is no prospect of monetary gain [7]. People can also be motivated to participate an activity by social and ethical reasons, such as socializing with others, glory, or recognition by others.

The sharing of personal data in MCS applications (e.g., citywide pollution monitoring) can raise significant privacy concerns, with information (e.g., location, point of interests) being sensitive and vulnerable to privacy attacks. To motivate user participation, it should explore new techniques to protect user privacy while allowing their devices to reliably contribute data.

### C. Dealing with Low-Quality Data

The involvement of human participation in community sensing brings forth certain uncertainties to MCS systems. For example, anonymous participants may send incorrect, low-quality or even fake data to a data center. Data contributed by different people may be redundant or inconsistent. The same sensor may sense the same event under different conditions (for example, sensing ambient noise when placing the mobile phone in a pocket or at hand). Therefore, *data selection* is often needed to improve data quality, and we should explore methods on fault filtering, quality estimation, expert contributor encouragement, etc.

The aim of MCS is to extract high-level intelligence from a large volume of user inputs. Regarding to the value of intelligence and its beneficiary, we can classify it into three distinct dimensions [11], namely, *user awareness*, *ambient awareness*, and *social awareness*.

- *User awareness* refers to the ability to understand personal contexts and behavioral patterns. Examples include human location, human activity, and daily routine patterns.
- *Ambient awareness* concerns status information on a particular space, which ranges from a small space to a citywide area. Examples include space semantics (i.e., the logical type of a place, such as a restaurant), ambient contexts, and traffic dynamics (e.g., traffic jams, hotspots).
- *Social awareness* goes beyond personal contexts and extends to group and community levels. The objective includes social interaction analysis (e.g., group detection, friendship prediction), social event detection, and so on.

### D. Heterogeneous, Cross-Space Data Mining

The mobile crowd data are collected from both offline and online communities. Different communities have distinct features in terms of geographical coverage, infrastructure support, function time, and so on. This also leads to distinct human interaction patterns (e.g., comment/like in online communities, co-location in ad hoc communities) and implicit social knowledge (e.g., friendship/trustworthy in online communities, social popularity/movement patterns in offline communities) that can be extracted from them. Study of the association and fusion of cross-space data (e.g., how does online social network data mirror physical events), as well as merging their complementary features and fully combining their merits (e.g., connecting the two forms of communities to enhance data transmission), however, become an important yet challenging research issue for MCS.

## IV. A REFERENCE FRAMEWORK

Based on the elaboration of MCS characters and applications, we propose a reference architecture to illustrate the key functional blocks and explain the key techniques of

MCS. It is intended to be the starting point that advances this new research area. Figure 2 shows the proposed architecture, which consists of five layers: *crowd sensing*, *data transmission*, *data collection*, *crowd data processing*, and *applications*.

*(1) Crowd sensing*. The first layer is a physical layer. Various everyday devices connect themselves to large networks. There are two data types generated from these devices: mobile sensing data and mobile social network data. The large-scale, raw data sensed should be shipped to the backend server for high-level intelligence extraction. *Access control* is another important function at the local side, where users can decide to whom her data can be shared.

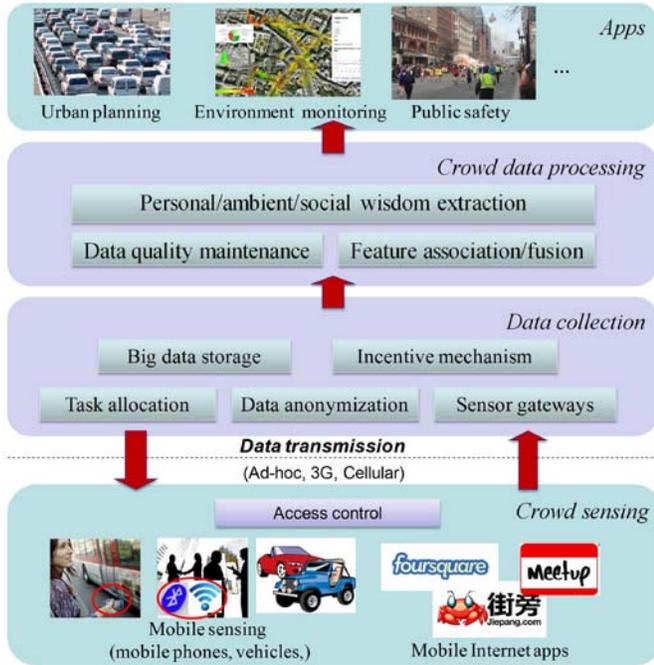

Figure 2. A reference framework for MCS.

*(2) Data transmission*. There are several mobile networking and communication techniques that can be leveraged by MCS, including ad hoc or opportunistic networks [12] (e.g., Bluetooth, Wi-Fi) and infrastructure-based networks (e.g., 3G, cellular). MCS applications should make data uploading transparent to the participant and tolerant of inevitable network interruptions.

*(3) Data collection infrastructure*. This layer gathers data from selected sensor nodes and provides privacy-preserving mechanisms for data contributors.

*(4) Crowd data processing*. This layer applies diverse machine learning and logic-based inference techniques to transform the collected low-level, single-modality sensing data into the expected intelligence. The focus is to mine the frequent data patterns to derive the three dimensions of crowd intelligence at an integrated level.

*(5) Applications*. This layer includes a variety of potential applications and services enabled by MCS. Associated functions include data visualization and user interface.

## V. THE HYBRID HUMAN-MACHINE INTELLIGENCE DESIGN IN MCS

The study of the combination of human and machine intelligence has a long history. Alan Turing wrote in 1950: "*The idea behind digital computers may be explained by saying that these machines are intended to carry out any operations which could be done by a human computer*" [13]. He also proposed a general procedure to test the intelligence of an agent now known as the Turing test. It represents that human intelligence and machine intelligence have been interlinked since the birth of AI research. Licklider's "man-computer symbiosis" [14] in 1960 also presents the idea that humans and computers can work together in complementary roles. MCS tries to solve the large-scale sensing/computing problems by having human in the loop. The reason is that human and machine intelligence show different strengths and weaknesses in MCS systems.

- *Human intelligence*. Knowledge, cognition, perception, and social interaction are general abilities of human beings. With these, human can have deep context and understanding of the sensing tasks. However, they are limited in memory and speed. Also, people vary in quality, and they often introduce errors or low quality data in MCS.
- *Machine intelligence*. Machines are powerful in large-scale storage and computing. Advanced data mining and machine learning algorithms also enable automatic knowledge discovery and event/society understanding. However, till now there are still numerous problems that cannot be addressed well by machines.

By combining computing with the intelligence of crowds (large groups of people participate in the sensing process), MCS allows the creation of hybrid human-machine systems. These hybrid systems enable applications and experiences that neither crowds nor computation could support alone. As far back as Ivan Sutherland's Sketchpad [15], *human-computer interaction has been structured around a tradeoff between user control and system automation*. The same is true for MCS, where we should investigate how to design the MCS system by mixing human and machine intelligence—a question that has not yet been solved by existing studies.

Figure 3 gives a description of our vision on the potential combination of human and machine intelligence in MCS systems. Over the three layers of MCS, human and machine intelligence can take complementary roles in terms of their distinct abilities. For example, in the data collection layer, people can understand and execute the tasks using their knowledge and cognition abilities. Machines, nevertheless, can decompose complex tasks and allocate them to proper human nodes; it further provides platforms for information sharing (e.g., user-contributed data in LBSNs).

There have been recently a few studies that try to combine the efforts of human and machine intelligence in MCS. DietSense [16] is a system that uses both automatic image processing techniques and manual image review, due to the complexity or ambiguity of recognition tasks. Kamar et al. showed how learned probabilistic models can be used to fuse human and machine contributions and to predict the behaviors of workers in online crowdsourcing [17].

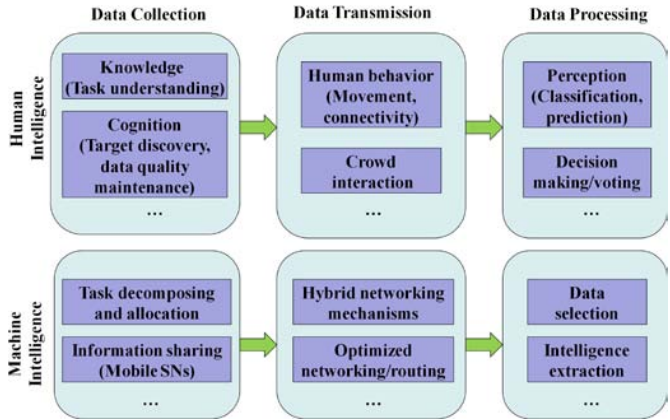

Figure 3. Complementary human-machine intelligence in MCS.

In addition to our vision in Fig. 3 and the above studies. We have two more suggestions on MCS design. First, the combination of human and machine intelligence should be application-centric. That is, we should create systems that dynamically trade-off human and machine intelligence in terms of application needs. Second, investigation of formal models and design patterns for crowd computing systems should be studied, which may make use of multi-disciplinary knowledge, including social science, management, computer science, and so on.

## VI. FUTURE RESEARCH TRENDS AND OUR EFFORTS

The study of MCS is still at its early stage and there are numerous challenges and research opportunities in the coming years.

*(1) A Generic Framework for Data Collection*. In MCS, mobile sensors from a highly volatile swarm of sensing nodes can potentially provide coverage where no static sensing infrastructure is available. Nevertheless, because of a potentially large population of mobile nodes, a sensing task must identify which node(s) may accept a task. A set of criteria should be considered in filtering irrelevant nodes: the specification of a required region (e.g., a particular street) and time window, acceptance conditions (for a traffic-condition capture task, only the phones out of user pockets and with good illumination conditions can satisfy requirements), and termination conditions (e.g., sampling period). Some preliminary studies on these issues have already been initiated. For example, in [18], a task description language called AnonyTL was proposed to specify the sample context for a sensing task. In [19], Reddy et al. developed a recruitment framework to enable organizers to identify well-suited participants for data collections based on geographic and temporal availability as well as participation habits. However, improving the efficiency of the decision making process in task assignment and data sampling necessitates further efforts.

*(2) Hybrid Mobile Networking*. As presented earlier, there are varied communication modules in mobile phones. The infrastructure-based connection uses pre-existing infrastructure (e.g., base stations, routers, access points) and manages the data in a centralized manner. The *ad hoc* or *opportunistic* connection, however, is founded on the development of opportunistic networks [12], which uses infrastructure-free, short-range radio techniques (Bluetooth, Wi-Fi, etc.) to build decentralized, ad hoc networks.

The two forms of networks have distinct merits and work environments. *Infrastructure-based networks* (e.g., cellular, 3G) can be accessed to people in the environments with Internet connection (e.g., at home, in the office, hot spots with wireless access points). *Opportunistic networks* are human-centric because they inherently follow the way that people opportunistically get into contact. For instance, customer *A* can connect with other customers that opportunistically meet in a coffee shop to build an ad-hoc mobile phone network. They have advantages on connecting and providing collaboration services to traveling users, who have difficulty in connecting the network infrastructure.

In the past years, significant research efforts have been made on facilitating data transmission/sharing in both infrastructure-based and opportunistic networks. However, they follow separate research lines, and the interlinking of the two forms of communities has little been explored. To address the transient network issue in MCS, it is important to explore the advantages of different networks and develop hybrid networking techniques. We have explored the hybrid networking protocol in our prior work — Hybrid Social Networking [20], which addresses the interlinking of online and opportunistic networking techniques to enhance data dissemination/transmission in MCS.

*(3) Varied Human Grouping*. Interaction among the volunteers is necessary, at least should be an option, but absent in most of current MCS systems. Researchers also advocated the need to facilitate interaction. Vukovic [21] claimed that one of the research challenges in crowd sensing is "*designing a mechanism for virtual team formation, incorporating not only skill-set, but also discovered social networks*". Lane [22] also believed that crowdsourcing misses automated methods to identify and characterize user communities. The interaction among users can also enhance the data quality in MCS. Therefore, grouping users and facilitate the interaction among them should be a challenge of MCS. Key techniques to address this include community creation approaches, dynamic group formation metrics, social networking methods, and so on. For example, to identify the people who are involved in the same social event, MoVi proposes a multi-dimensional sensing approach, where a combination of visual and acoustic ambience of phones are used [23]. In GroupMe [24], we also tackled how to extract and suggest groups using mobile phone sensing techniques.

*(4) Cross-Community Sensing and Mining*. Data from different communities often present different attributes and strengths, and moreover, they are often complementary. MCS explores the integration of data from both online (e.g., mobile social networks) and offline (e.g., mobile sensing) communities to demonstrate their aggregated power in various purposes, such as the enrichment of trip planning mentioned in the introduction. Here, we describe two more examples to showcase the effects of data integration from distinct communities.

- *Sensor-based activity recognition enhanced by Web-mined knowledge.* Knowledge obtained from the Web can be used to assist activity recognition in the physical world. For instance, Philipose et al. extracted the activity-relevant objects from the Web (Wikis, HowtoDos), which is then used in RFID-based human activity recognition [25].
- *Merging the data from heterogeneous communities to develop new social apps.* Data from different spaces often characterizes one facet of a situation, thus the fusion of data sources often draws a better picture of the situation. For example, by integrating the mined theme from user posts and the revealed location information from GPS-equipped mobile phones, Twitter has been exploited to support near real-time report of earthquakes in Japan [26].

We have also organized a special issue titled "cross-community mining" [27] on the *Personal and Ubiquitous Computing journal*, where more examples to leverage data from distinct communities can be found.

## VII. CONCLUSION

In this paper we have presented Mobile Crowd Sensing (MCS). Similar to participatory sensing, MCS also leverages the power of citizens for large-scale sensing. However, it goes beyond participatory sensing by having implicit and explicit participation, and the collection of crowdsourced data from both mobile sensing and mobile social network services. We have characterized the key features of MCS, including crowd-powered data collection, cross-space data mining, and low quality data analysis. To facilitate the development of MCS apps, we propose a reference framework and discuss the efforts of balancing human and machine intelligence in MCS system design. As an emerging research area, MCS brings numerous issues to be addressed, such as hybrid networking, varied user grouping, cross-community sensing and mining, and so on. These new challenges will bring unprecedented opportunities to academic researchers, industrial designers/developers, as well as policy makers.

## ACKNOWLEDGEMENTS

This work was partially supported by the National Basic Research Program of China 973 (No. 2012CB316400), the National Natural Science Foundation of China (No. 61332005, 61373119, 61222209, 61103063), the Natural Science Basic Research Plan in Shaanxi Province of China (No. 2012JQ8028).